# Single-pixel digital "ghost" holography

Pere Clemente[1], Vicente Durán[1,2], Enrique Tajahuerce[1,2], Victor Torres-Company[3*] and Jesus Lancis[1,2]

[1]Institut de Noves Tecnologies de la Imatge (INIT), Universitat Jaume I, E12080 Castelló, Spain

[2]GROC·UJI, Departament de Física, Universitat Jaume I, E12071 Castelló, Spain

[3]Microtechnology and Nanoscience department (MC2), Chalmers University of Technology, SE-41296, Gothenburg, Sweden

[*]torresv@chalmers.se

Since its discovery, the "ghost" diffraction phenomenon has emerged as a non-conventional technique for optical imaging with very promising advantages. However, extracting intensity and phase information of a structured and realistic object remains a challenge. Here, we show that a "ghost" hologram can be recorded with a single-pixel configuration by adapting concepts from standard digital holography. The presented homodyne scheme enables phase imaging with nanometric depth resolution, three-dimensional focusing ability, and shows high signal-to-noise ratio.

**Introduction.-** In 1948 Gabor introduced the technique of holography to push the resolution of electron microscopy near to its theoretical limit [1]. The invention of the laser boosted the field and the first practical 3D holograms were achieved [2]. Recording interference patterns with a charge-coupled device (CCD) led to the development of digital holography, where the intensity and phase of the electromagnetic field is measured, stored, transmitted and manipulated with the aid of a computer [3]. Digital holography is currently a ubiquitous diagnostic and metrology tool [4, 5]. Taking biological applications as an example, digital holographic microscopy offers the capability to measure phase variations in the nanometer range, allowing marker-free quantitative analysis in the cellular and sub-cellular range [6]. Nowadays, digital holographic microscopes are so compact and versatile that can be integrated in a cell-phone to offer a cost-effective tool for telemedicine applications [7].

In a parallel research avenue, the so-called "ghost" imaging technique continues attracting attention since its demonstration in the mid 90's [8- 11]. It permits to use single-pixel detectors [12], offers enhanced robustness to weakly absorbing samples [13], features the cancellation of optical aberrations [14], and can be used for image encryption [15]. In its most standard configuration, two light beams from a common light source propagate through different optical systems. The light illuminating the object is collected with a detector without spatial resolution (so-called "bucket"). The term "ghost" comes from the counterintuitive fact that the image appears by correlating the intensity distributions obtained at the path where the object was not located with the outcomes from the bucket detector. The early experiments used spontaneous parametric downconversion (SPDC) as light source, which may show entanglement. Later, it was proved that many features obtained in a "ghost" imaging experiment could also be reproduced with a pseudo-thermal classical light source [16]. After two decades of intense debates, we now understand that it can be implemented with either quantum or classical sources, albeit with slightly distinct features [16-21]. Leaving aside the quantum vs. classical debate, the "ghost" appellative actually offers enormous possibilities for optical imaging hitherto not fully exploited. In a recently proposed configuration, a laser is used as light source and the spatial incoherence is introduced by a set of random phase patterns pre-programmed on a spatial light modulator (SLM) [20]. In this way, only the beam path corresponding to the single-pixel detector needs to be in the setup, as the propagation through the other arm is calculated offline. This hybrid digital-optical scheme enables three-dimensional sectioning while simply using a point detector [12]. The computation comes at the expense of storing and processing thousands of speckle patterns to achieve an image with suitable signal-to-noise ratio (SNR), a problem that can be partly alleviated by using compressive sensing algorithms [22].

Advanced imaging applications require amplitude and phase information of structured objects, which represents a significant challenge for any "ghost" imaging setup. A few attempts can be found in the literature. Using 2D phase-retrieval algorithms, the object's complex information can be extracted from the near- and far-field patterns [23,24], but only after re-arranging the optical setup for each configuration. Another approach corresponds to Ref. [25] where the complex information is extracted from the measurement of the first-order spatial cross-correlation function using a modified Young's interferometer. Both configurations are rather cumbersome and still need a 2D sensor. Finally, quantum entanglement was also proposed for extracting holographic information about a remote 3-D object in a confined space, although the experimental verification was not implemented [26]. Here, we show a homodyne configuration that allows for recording the "ghost" Fourier digital hologram of an object in a hybrid digital-optical scheme while using a single-pixel detector. This scheme provides an inherently higher

SNR than other intensity-based configurations. Furthermore, our solution allows for 3D sectioning and full-frame phase imaging with nanometer resolution of highly structured samples.

**Theory.-** Our configuration bears intriguing similarities with the conventional phase-shifting digital holography (PSDH) technique [27]. A comparison between both schemes is shown in Fig. 1. Each setup consists of a Mach-Zehnder interferometer, which includes a phase shifter in one of the arms to introduce stepwise phase differences $\Delta\varphi$. In PSDH, at least three phase-shifted interferograms are recorded with a CCD [27], as shown in Fig. 1(a). From these intensity patterns, the complex amplitude of the object is reconstructed offline with a proper algorithm. In our digital "ghost" holography (DGH) setup, spatially coherent light from a monochromatic laser is split in two arms. In the object arm, we insert the sample whose complex information is to be retrieved and an SLM to shape the phase distribution of the impinging beam. The object wave is recombined with the reference beam and the central point of the interference pattern is measured with a pinhole detector. It is important to recognize that, now, the irradiance distribution at the output plane is sampled at a single point, which is the major difference with conventional PSDH. The role of the SLM has a twofold aim. First, as in computational "ghost" imaging, it imprints a set of pre-established random phase distributions to generate speckle patterns at the object plane. Second, on each distribution it introduces sequentially a set of constant phase-shifts, which are mandatory for complex field reconstruction.

The optical field $O_{i,\varphi}(\vec{r}_1)$ at the detection plane, with transversal coordinate $\vec{r}_1$ in Fig. 1(b), can be calculated from elementary paraxial diffraction theory,

$$O_{i,\varphi}(\vec{r}_1) = \left|\left\{\left[\exp\left[j(\Phi_i(\vec{r}_1) + \varphi)\right] \otimes h_{z'-z}(\vec{r}_1)\right] t(\vec{r}_1)\right\} \otimes h_z(\vec{r}_1) + u(\vec{r}_1)\right|^2, \qquad (1)$$

where $\Phi_i(\vec{r})$ is the *i*th random phase distribution codified onto the SLM, $h_z(\vec{r})$ is the Fresnel kernel propagator, given by $h_z(\vec{r}) \propto \exp[jk\vec{r}^2/(2z)]$ (with *k* the wavenumber), $t[\vec{r}=(x,y)]$ is the object's complex amplitude and $u(\vec{r})$ the unshaped laser field distribution (in our experiment considered to be a plane wave). The symbol $\otimes$ stands for convolution operation, and *z* and *z'* are the distances to the detector from the object and from the SLM, respectively. The above equation accounts for the paraxial propagation from the SLM to the object plane, transmission through the mask, and further propagation till the detector, where it interferes with the reference wave. A constant phase-shift $\varphi$, both 0 and $\pi$, is added to the phase programmed onto the SLM a sequential way. For each value of the phase-shift, the signal at the output plane is sampled by a single-pixel detector, which is arbitrarily chosen to be at the origin. In mathematical terms, we measured the signal $B_{i,\varphi} = O_{i,\varphi}(\vec{r}_1 = \vec{0})$. The above operation is repeated for *N* realizations, and the weight coefficient $\Delta B_i = (B_{i,0} - B_{i,\pi})$ is measured at each one of them.

Additionally, we compute offline the 2D interference pattern at the output plane with transversal coordinate $\vec{r}_2$ when the object is removed. The intensity is given by

$$I_{i,\varphi}(\vec{r}_2) = \left|\exp[j(\Phi_i(\vec{r}_2)+\varphi)] \otimes h_{z'}(\vec{r}_2) + u(\vec{r}_2)\right|^2. \qquad (2)$$

In Eq. (2), we have used the following property of the Fresnel kernel $h_{z-z'}(\vec{r}) \otimes h_{z'}(\vec{r}) = h_z(\vec{r})$. For the computational part of the experiment, three different values for the phase shift must be considered $\varphi = (0, \pi/2, \pi)$ and the signal

$$\Delta I_i(\vec{r}_2) = \left[I_{i,0}(\vec{r}_2) - I_{i,\pi}(\vec{r}_2)\right] - j\left[2I_{i,\pi/2}(\vec{r}_2) - I_{i,0}(\vec{r}_2) - I_{i,\pi}(\vec{r}_2)\right] \qquad (3)$$

is retained. Note that this quantity can be calculated as long as the random phase $\Phi_i(\vec{r})$ is known, which is a major difference with regard to non-computational approaches to ghost imaging experiments.

Finally, the object's information is reconstructed from the correlation between a set of measured intensities $\{\Delta B\}$ and a set of computed intensity patterns $\{\Delta I\}$,

$$G(\vec{r}_1=0,\vec{r}_2) = \langle \Delta B \Delta I(\vec{r}_2)\rangle, \tag{4}$$

where the brackets denote ensemble average. In practical terms, the above equation is calculated as $G(\vec{r}_1=0,\vec{r}_2) = 1/N \sum_i^N \Delta B_i \Delta I_i(\vec{r}_2)$. The symbol $\Delta$ points out that the quantities to be correlated are intensity differences calculated with the aid of Eqs. (1) and (3). After a cumbersome but straightforward calculation, we get

$$T(\vec{u}) \propto \exp\left(j\frac{k}{2z}|\vec{r}_2|^2\right) G(\vec{r}_1=0,\vec{r}_2). \tag{5}$$

Here, $T(\vec{u})$ denotes the Fourier transform of the complex object $t(\vec{r})$, with the spatial frequency $\vec{u} = \vec{r}/\lambda z$, where $\lambda$ is the wavelength of the laser. The above equation indicates that, through our new ghost holography approach, it is possible to retrieve the amplitude and phase information of a complex sample.

**Experiment**.- The experimental setup was based on the scheme shown in Fig 1(b). We used a He-Ne laser emitting at 0.6328 $\mu$m as the spatially coherent light source. The SLM was a reflective 2D liquid crystal on silicon display (LCoS-SLM, Holoeye LC-R 2500) with XGA resolution and a pixel pitch $\delta x_0$ of 19 $\mu$m. The random phase patterns $\Phi_i(\vec{r})$ were generated by uploading a sequence of $N = 24000$ random images of 256×256 pixels onto the modulator panel.

The random phases were uniformly distributed in [0, 2π] radians, and each phase value was realized by group clusters of 2×2 SLM pixels. The refreshing frequency of the LCoS display, which fixed the maximum system speed, was 60 Hz. As a pinhole detector we used a 6.45 μm-sized pixel of a commercial CCD (Basler A102f).

Figure 2 shows the object reconstruction from the "ghost" Fourier hologram of an amplitude object (a capital letter of 2×2.7 mm). As expected, the object's phase (expressed in radians) is approximately uniform (with a relative standard deviation up to 2 % due to glass roughness). The transverse resolution of the "ghost" digital Fourier hologram is given by the size $\delta x$ of the speckle grains generated by the modulator on the detection plane. For a window of $m \times m$ pixels displayed on the SLM plane, $\delta x = 0.9 \lambda z'/m\delta x_0$ [28]. For the results shown in Fig. 2, $z' = 80.5$ cm and thus $\delta x \cong 95$ μm.

A major challenge for "ghost" imaging is to retrieve the object's phase distribution. Recent proposals for phase-only imaging include non-local filtering using pseudothermal radiation and 2D image detection and correlation [29]. However, phase retrieval in holographic measurements is quantitative and depth resolution of the phase sample is of the order of a fraction of the wavelength, i.e., is in the range of the nanometer for visible light. Experimental results for a transparent plate that generates a Seidel aberration (the Zernike polynomial $Z_3^1$) are presented in Fig 3(a). Phase values are obtained from the reconstruction of the "ghost" digital Fourier hologram using the FFT algorithm. The maximum phase value of the reconstructed pattern corresponds to an optical path difference of ~8$\lambda$. In the zones where the phase gradient is higher, clearly visible in the wrapped pattern, the phase gradient reaches values up to 0.8 rad/pixel (i.e., around 75 nm in axial optical path length).

Our configuration also enables zooming capabilities with minimum reconfiguration by adjusting the distances $z$ and $z'$. Figure 3 (a) (down) illustrates this ability. This image represents a 1.6x resolution improvement when compared to the upper Fig. 3(a), at the expense of a corresponding reduction in spatial aperture. Finally, the fact that the intensity and phase information is recovered allows for 3D wavefield reconstruction through digital forward-propagation, a feature illustrated in Fig. 3(b). In other words, we can calculate the diffracted light pattern generated by the sample in any transverse plane from the registered hologram light distribution. The results obtained with single-pixel DGH are in good agreement with those obtained using a CCD (Basler A102f) in the conventional PSDH setup shown in Fig 1(a).

It is well known that homodyne detection provides imagery recovery with higher SNR [30, 31]. Our proposal provides an added advantage, namely, it works with a single pixel. In Fig. 4 we simulate the SNR versus the number $N$ of computed speckle patterns for the binary mask in Fig. 2. Results of intensity reconstruction of the "ghost" Fourier hologram are compared with those obtained from computational "ghost" imaging through the algorithm detailed in Ref. [12]. The parameters used in DGH calculations are the same as those considered in Fig. 2 and the SNR is calculated following the procedure reported in [22]. The SNR of the recovered images with DGH increases proportional to $N$, in contrast to $\sqrt{N}$ as in conventional "ghost" imaging [11, 12].

**Conclusions.-** We have reported a "ghost" scheme that measures the Fourier hologram of a sample object. This has allowed us to get the complex information of structured and realistic objects with high SNR, a hitherto challenge for any ghost imaging setup. Appealing enough, our scheme only uses a single-pixel detector in the physical configuration and yet recovers 3D information of an object. We expect these results will bring phase-shifting digital holography technologies at other spectral regions (like the terahertz or infrared), where 2D sensor displays are costly or simply not available.

**Acknowledgments**

This work has been partly funded by the Spanish Ministry of Education (project FIS2010-15746) and the Excellence Net from the Generalitat Valenciana about Medical Imaging (project ISIC/2012/013) and through Prometeo Excellence Programme (project PROMETEO/2012/021).

# Figures

Figure 1 (color online)**.** Schematic comparison of (a) phase-shifting digital holography and (b) the proposed digital "ghost" holography.

(a)

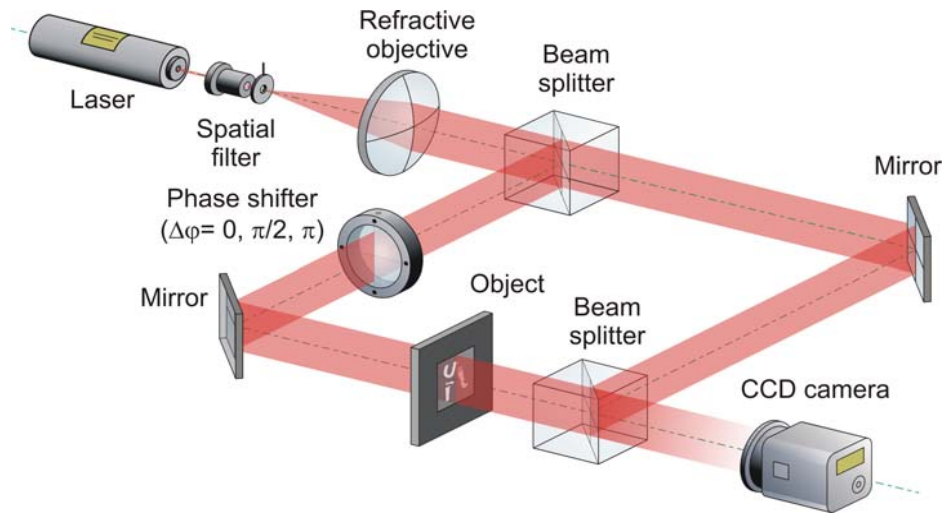

(b)

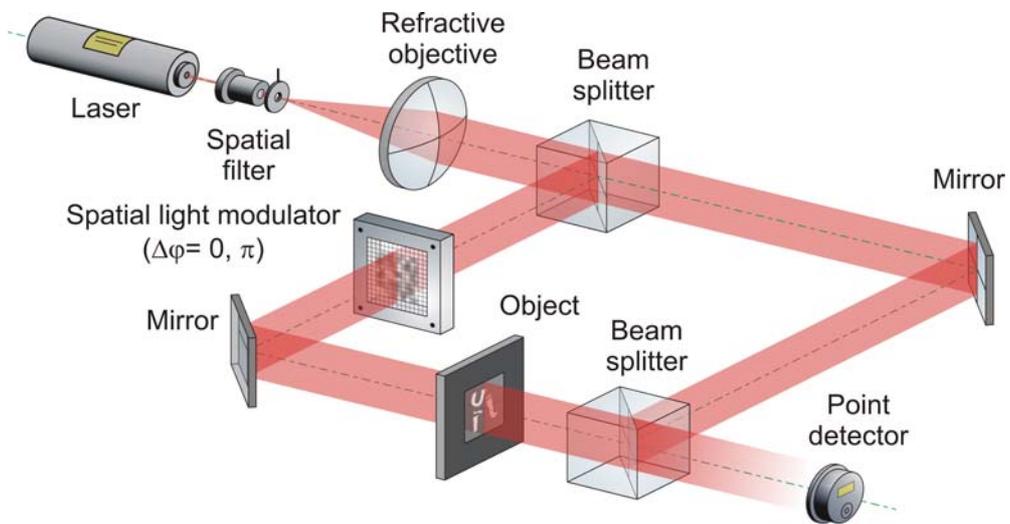

Figure 2 (color online). Amplitude and phase information recovered from the Fourier "ghost" hologram of an amplitude mask.

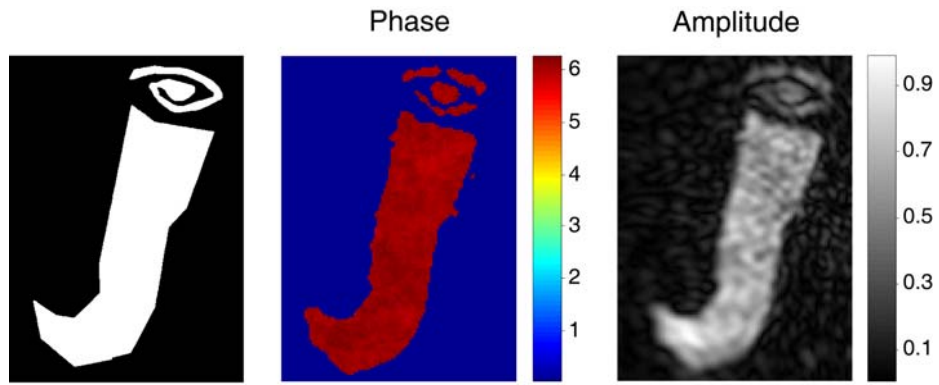

Figure 3 (color online). "Ghost" holography results for a phase-only object, an aberrated circular pupil inscribed in a square of 8×8 mm$^2$. (a) Up. Wrapped and unwrapped phase patterns (in radians). The setup distances are $z = 63$ cm and $z'=81$ cm. Down. Zooming capability of the digital "ghost" holography scheme (with $z = 40$ cm and $z'=80.5$ cm). (b) Comparison of 3D wavefield reconstruction from the recovered phase pattern with DGH and PSDH.

(a)

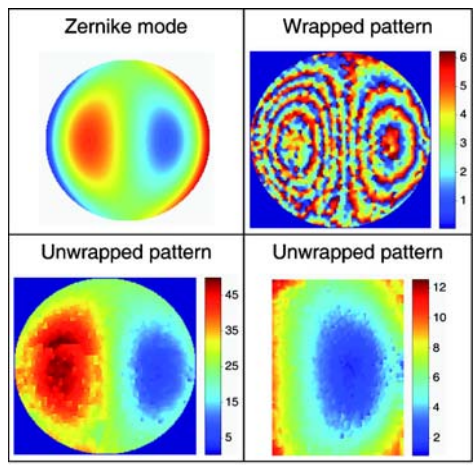

(b)

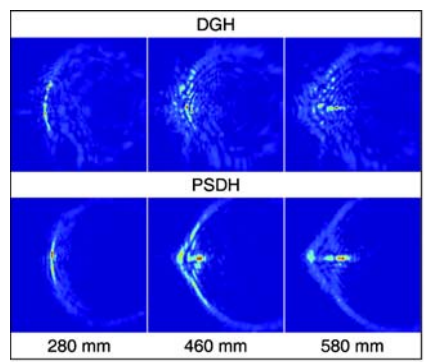

Figure 4 (color online): Evolution of SNR vs. number of computed speckle patterns for the binary mask of Fig. 2. The upper curve corresponds to DGH.

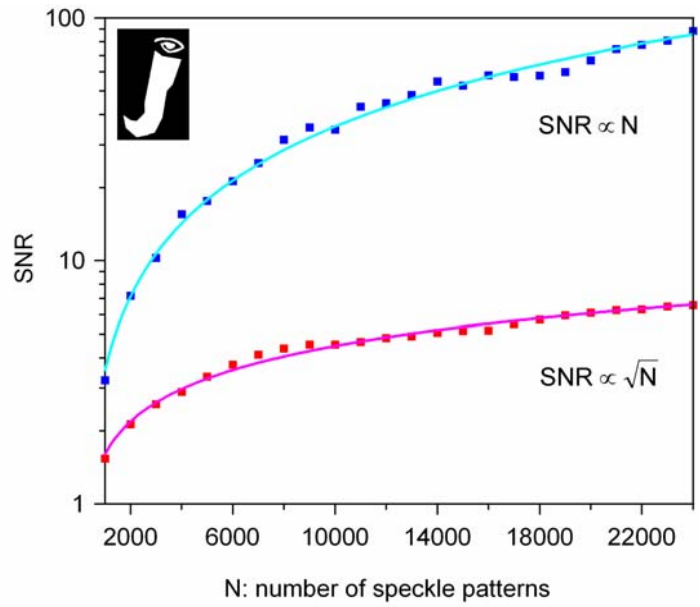